# Isotope engineering of silicon and diamond for quantum computing and sensing applications

Kohei M. Itoh[*1] and Hideyuki Watanabe[*2]

[*1] School of Fundamental Science and Technology, Keio University, Yokohama 223-8522, Japan

[*2] Electronics and Photonics Research Institute, National Institute of Advanced Industrial Science and Technology (AIST), Tsukuba 305-8562, Japan

Corresponding author: Kohei M. Itoh, kitoh@appi.keio.ac.jp

Abstract

Some of the stable isotopes of silicon and carbon have zero nuclear spin, whereas many of the other elements that constitute semiconductors consist entirely of stable isotopes that have nuclear spins. Silicon and diamond crystals composed of nuclear-spin-free stable isotopes ($^{28}$Si, $^{30}$Si, or $^{12}$C) are considered to be ideal host matrixes to place spin quantum bits (qubits) for quntum computing and sensing applications because their coherent properties are not disrupted thanks to the absence of host nuclear spins. The present article describes the state-of-the-art and future perspective of silicon and diamond isotope engineering for development of quantum information processing devices.

## I. INTRODUCTION

Quantum computing and sensing are emerging concepts that can surpass the theoretical achievable limit of current classical computing and sensing technologies. The realization of quantum computers and sensors in industrially compatible solid-state platforms such as silicon and diamond will allow their integration with state-of-the-art classical devices such as silicon integrated circuits, compound semiconductor lasers and photodiodes, metallic magnetic mediums, and micro electro-mechanical systems. This article describes the present status and future of silicon quantum computing and diamond quantum sensing research, from the point of view of materials scientists who have been developing the isotope engineering of silicon and diamond. While isotope engineering has been employed over the past two decades to understand the basic properties of semiconductors,[1, 2] it was B. E. Kane who foresaw its importance in quantum computation applications in 1998.[3] Here, the use of the single $^{31}$P nuclear spin in each phosphorus donor placed in silicon as a quantum bit (qubit) was introduced. The importance of eliminating background $^{29}$Si nuclear spins (4.7% isotopic abundance in naturally available silicon) was suggested because $^{29}$Si nuclear spins could act as a source of magnetic noise, disturbing the $^{31}$P nuclear spin quantum information.[3] Such elimination of background host nuclear spins is not possible with widely studied III-V semiconductor quantum structures. Other researchers, including one of the present authors, have suggested that $^{29}$Si nuclear spins could be employed as qubits if their positions can be controlled at the atomic level in a $^{28}$Si host.[4-6] At around the same time, the nitrogen-vacancy (NV) center in diamond was identified as a single qubit that could be operational at room temperature.[7-9] This discovery was followed immediately by studies to use NV qubits for quantum computing[10] and sensing (metrology).[11-14] Similar to the case of silicon, the enrichment of diamond by the nuclear-spin-free $^{12}$C stable isotope preserves the quantum information in the NV centers.[15] This review provides a perspective on the use of isotope engineering for quantum computation and sensing using silicon and diamond. For general aspects of the challenges associated with materials science, readers are encouraged refer to October's 2013 issue of the MRS Bulletin featuring *materials issues for quantum computation*.

## II. SILICON QUANTUM COMPUTATION

### A. SILICON-BASED QUANTUM COMPUTER PROPOSALS

A review on the developments of silicon-based quantum computers was published recently.[16] This subsection provides a brief summary on the studies on Si-based quantum computers that require isotope engineering. Figure 1 shows two examples of silicon quantum computer proposals that use isotope engineering. The report by Kane (Fig. 1(a)) proposed to use the nuclear spins of phosphorus donors embedded in isotopically enriched $^{28}$Si as qubits.[3] The phosphorus donors form a one-dimensional array with a separation of ~20 nm. The quantum information stored and processed by each $^{31}$P nuclear spin (I=1/2) is read out by the tunneling current induced by the electron bound to $^{31}$P as it transports to the adjacent phosphorus donor. At the beginning of the calculation, each nuclear spin is initialized so that they have identical nuclear spin states, $m_s$=1/2 or −1/2. This can be achieved by reading out each nuclear spin state using the tunneling current and performing nuclear magnetic resonance (NMR) on the $^{31}$P qubits to align them into the correct state when their initial state is incorrect. No action is taken if the $^{31}$P qubit is already in the correct state. NMR is also used for the quantum logic operation of the $^{31}$P qubits. Qubit selectivity is achieved by applying an electronic bias to a gate that is placed immediately above each qubit (A-gate). This causes the $^{31}$P to be either in or out of resonance with the frequency employed during the NMR. The two-qubit interactions are switched on and off by electric gates placed between the neighboring qubits (J-gates). The electrons bound to the phosphorus donors mediate the $^{31}$P-$^{31}$P nuclear spin interactions between the adjacent qubits. The nuclear spin state for each $^{31}$P qubit is read out at the end of the calculation by detecting the single-electron tunneling current using a single-electron transistor (SET). Kane's vision has triggered many proposals that use the electron spins in donors imbedded in silicon as qubits.[17-19]

The second class of silicon-based quantum computation studies uses $^{29}$Si nuclear spins as qubits.[4-6] Figure 1(b) shows the one-dimensional array of the $^{29}$Si nuclear spins formed on an isotopically enriched nuclear-spin-free $^{28}$Si substrate.[6] Here, the dipolar coupling between adjacent $^{29}$Si nuclear spins is used for two-qubit operations and the qubit selectivity is ensured by the large magnetic field gradient induced by a small magnet placed near the array. Readout of each qubit is achieved by shuttling the state to the end of the chain by a sequence of swap operations between the adjacent qubits. The $^{29}$Si nuclear spin state at the end of the chain was read out from the electron in the phosphorus atom that is placed next to the $^{29}$Si atom.

The third class of silicon-based quantum computing studies uses the electrons in quantum dots as qubits.[20-22] A single-electron spin is confined in each quantum dot and is employed as a qubit. The qubit state is described by either the spin state of electrons in a dot or a single electron within a pair of two quantum dots.[16]

## B. ISOTOPICALLY ENGINEERED SILICON FOR QUANTUM COMPUTATION

Naturally available silicon ($^{nat}$Si) is composed of 92.2% $^{28}$Si, 4.7% $^{29}$Si, and 3.1% $^{30}$Si. These stable isotopes can be separated by centrifuging SiF$_4$ gas.[23, 24] The first isotopically engineered silicon was produced in 1958 to characterize the electron spin resonance of donors in silicon.[25] Since then, no results on this topic were reported until a strong interest emerged in the mid-1990s, when isotopically enriched $^{28}$Si single crystals were employed as the standard to define the Avogadro constant.[26-28] The goal was to produce silicon crystals with a well-defined number of Si atoms per unit volume and to establish the definition of the kilogram by the mass of a given number of silicon atoms. Silicon was chosen because there had been extensive industrial efforts in Si electronics. Silicon has unsurpassed crystalline quality and chemical purity in comparison with other crystalline materials. Isotope enrichment with the stable isotopes is needed because the Si-Si bond length depends on the combination of isotopes.[29, 30] Fluctuations in the bond lengths for different isotopes must be removed to obtain a well-defined number of Si atoms per unit volume. The floating-zone growth performed at the Leibniz Institute for Crystal Growth (IKZ) in Berlin resulted in a 5-kg $^{28}$Si single crystal of 99.99% isotopically enriched silicon with <10$^{15}$ cm$^{-3}$ background carbon content, <10$^{13}$ cm$^{-3}$ electrically active impurities such as boron and phosphorus, and was free of dislocations.[31, 32] The silicon crystals fabricated for the Avogadro measurement were shaped into precise spheres and the sawn-off pieces generated during the shaping process were used to perform basic characterization on the crystal. Some pieces were also used in some of the silicon quantum information research described below. Herein, these samples will be referred to as the "Avogadro samples". They had the highest isotopic and chemical purity and the best crystalline quality among all of the isotopically enriched silicon crystals ever produced.

The second group of isotopically enriched, bulk Si crystals was also grown at the IKZ and will be referred to as the "IKZ crystals". IKZ established various Czochralski[23, 33] and floating-zone[34] growth techniques to maintain the isotopic composition of the starting charge. Here various isotopes ($^{28}$Si, $^{29}$Si, and $^{30}$Si) of different enrichements were employed as stratting charges as described in Ref. 23. They also developed a method to dope the crystal with the desired group V donors during its growth. A variety of dislocation-free, isotopically enriched $^{28}$Si, $^{29}$Si, and $^{30}$Si crystals were grown at IKZ to fulfill orders placed by the Simon Fraser University, Canada,[35, 36] the Keio University, Japan,[23, 37] and the University of California (UC), Berkeley, USA.[34, 38]

The third group of isotopically enriched, bulk Si crystals was grown at Keio University, Japan and will be referred to as the "Keio crystals". The floating-zone growth machine at Keio was designed to grow small crystals (<7 mm diameter) of high crystalline quality and chemical purity.[39] The machine is suited to grow isotopically mixed crystals with arbitrary compositions of $^{28}$Si, $^{29}$Si, and $^{30}$Si. Isotopically mixed bulk Si crystals have been produced exclusively at Keio University.[40]

The isotopically enriched $^{28}$Si chemical vapor deposition (CVD) films were grown from isotopically enriched silane ($^{28}$SiH$_4$) gas by the Isonics Corporation, USA, from 1998 to 2005. Because centrifuge isotope separation was performed using SiF$_4$ as described elsewhere [23], the resulting isotopically enriched $^{28}$SiF$_4$ was transformed to $^{28}$SiH$_4$ (silane) by an electronic materials company named Voltaix. The quality of the Isonics epilayers, especially the residual impurity concentration, fluctuated significantly between the different epilayers because the company outsourced the CVD growth to other companies. Therefore, obtaining quantum-information-research-grade $^{28}$Si CVD wafers, including $^{28}$Si on insulator (SOI) wafers,[41] required in-depth discussions with Isonics to establish appropriate fabrication procedures. High quality, Isonics $^{28}$Si ep-wafers employed in single qubit experiments (Section D) have been grown at Lawrence Semiconductor Research Laboratory, Inc.

Isotopically enriched, strained $^{28}$Si thin layers were grown by solid-source molecular beam epitaxy (MBE) at the Technical University of Munich (TUM), Germany[42] and by CVD at Princeton University, USA.[43] Some electron double quantum dots were fabricated using the strained silicon.[44] Other isotopically controlled silicon low-dimensional structures, typically grown by MBE,[45-48] have been employed, predominantly for diffusion[49-52] and amorphization[53] studies.

## C. PROOF-OF-CONCEPT EXPERIMENTS WITH SPIN ENSEMBLES IN ISOTOPICALLY ENGINEERED SILICON

A qubit must be made of a single quantum. However, one can prepare an ensemble of identical qubits, initialize them in the same quantum states, manipulate them together quantum mechanically (perform quantum calculations), and read them out all at once to obtain the results expected for single quantum bit operation. To demonstrate this, phosphorus donors in silicon were used. Figure 2(a) shows two phosphorus donors placed in $^{nat}$Si. Each phosphorus atom has a electron spin for the bound electron and the $^{31}$P nuclear spin in the nucleus; it has two qubits. This was used to determine whether the two phosphorus donors in Fig. 2(a) are quantum mechanically identical. They are not because of the coexistence of $^{28}$Si, $^{29}$Si, and $^{30}$Si stable isotopes and their random distribution in $^{nat}$Si, and the silicon isotope configuration around each phosphorus donor differed from those of others. Additionally, the mass difference between the stable isotopes changes the bond lengths (the local lattice strain), leading to perturbation of the hyperfine interactions between the electron spin and the $^{31}$P nuclear spin in each phosphorus atom. This could lead to a shift in the NMR frequencies of the $^{31}$P between different donors, making each phosphorus atom distinguishable and the ensemble of phosphorus donors would not be able to be manipulated all together. Moreover, there was a random distribution of $^{29}$Si that changes the local magnetic field near each phosphorus donor. This leads to even larger changes in the $^{31}$P NMR and electron paramagnetic resonance (EPR) frequencies than those caused by the mass fluctuations. Therefore, the silicon matrix must be composed of a single type of stable isotopes with no nuclear spins, either $^{28}$Si or $^{30}$Si. Figure 2(b) shows two phosphorus donors placed in an isotopically enriched, $^{28}$Si single crystal. The two phosphorus donors could be considered identical if the separation between them was long enough to have small interactions. This changes their EPR frequencies, where their quantum states could be initialized to be the same such that the electron and nuclear spins were set in the up (or down) state in every donor. By using an ensemble of identical phosphorus donors, one can start from the same quantum states, perform the same quantum calculation using every donor in parallel and read them all out together to obtain the result from the quantum calculation that is expected for a single phosphorus donor qubit. Therefore, isotope enrichment is a powerful and necessary tool to allow for proof-of-concept quantum computer experiments with an ensemble of identical qubits in silicon to be carried out.

One important issue yet to be mentioned thus far is the coherence time ($T_2$) of the qubits. If the coherence time (the quantum information storage time) is too short, particularly shorter than the total time required to complete the desired quantum computation routine, the quantum computer will lose information in the middle of the calculation. Because a high crystalline quality and low concentration of residual impurities can be achieved with silicon, the major decoherence source of the electron and nuclear spins in silicon has been assumed to be caused by magnetic field fluctuations arising from the flip-flops of the $^{29}$Si nuclear spins in the matrix.[54] The mass variations in the isotopes also change the local lattice strain to vary the spin manipulation frequencies. However, they are static and have little effect on the coherence time. Therefore, the first step was to understand the decoherence mechanisms of the electron and nuclear spins in silicon as a function of the host $^{29}$Si contents. An example of this is a series of pulsed EPR measurements performed on an ensemble of phosphorus donors using the Keio bulk crystals in which the $^{29}$Si fraction was deliberately changed.[40, 55-57] The spin coherence time of the phosphorus bound electron and its interaction with the precessing $^{29}$Si nuclear spins in the neighboring atoms [electron spin echo envelope modulation (ESEEM)] was determined experimentally and compared with the theory[58-60] to achieve a solid understanding of the decoherence mechanisms. The most important issue associated with quantum computation is how far the phosphorus coherence time can be extended by eliminating $^{29}$Si from the matrix. The first experiment of this type was performed by Gordon and Bowers in 1958 using an isotopically enriched, $^{28}$Si bulk single crystal. However, no significant improvement in the coherence was obtained because it was limited by the interactions between the donor electrons, caused by the relatively high concentration of phosphorus.[25] More recently, in 2003, an extended coherence time of the electron spins bound to the phosphorus donors in isotopically enriched, Isonics $^{28}$Si epilayers was reported by the Princeton University group.[61] The coherence was improved even further in 2011 by the international collaboration headed by the same group using the IKZ $^{28}$Si bulk crystals.[62] The study achieved electron coherence times exceeding seconds. This time is long enough to perform a quantum error correction protocol that, in principle, can store information forever. The dominant decoherence mechanisms were recorded as a function of temperature.[62] Because the $^{31}$P nuclear spins couple to the electron spins bound to phosphorus, the nuclear spin coherence can be affected by the presence of $^{29}$Si through hyperfine interactions with the electron spins. Therefore, the research team headed by the Simon Fraser University used the isotopically enriched, $^{28}$Si Avogadro crystal to probe the coherence time of the $^{31}$P nuclear spins in neutral[63] and ionized[64] donors. Thanks to the removal of the background $^{29}$Si nuclear spins, coherence time of the $^{31}$P nuclear spin in its neutral state at the cryonegic temperautere exceeded 180 s [63] while that of ionized donors at room and cryogenic temperatures exceeded 39 min and 3 hours, respectively.[64] The important finding here was that the coherence times of not only the electron but also the nuclear spins of the phosphorus donors in silicon were extended by eliminating the $^{29}$Si nuclear spins through isotope engineering. These coherence times are more than sufficient for practical application in quantum computing.

Following the success of phosphorus in isotopically enriched $^{28}$Si, the group headed by the Simon Fraser University and University College London investigated other shallow donors, focusing on bismuth in isotopically enriched $^{28}$Si.[65, 66] The coherence times of the bismuth electron spins were as long as those of the phosphorus atoms.[65] The existence of the optimal magnetic field (atomic clock transition), at which the electron decoherence was suppressed by its insensitivity to the external magnetic field perturbation was demonstrated.[66] The Keio University group investigated the hyperfine clock transition of bismuth using the magnetic field at which the resonant frequency was insensitive to fluctuations in the hyperfine constant in isotopically enriched $^{28}$Si[67] using the spin-dependent-recombination EPR technique developed especially for this purpose.[68]

The decoherence and decoupling mechanisms for the $^{29}$Si nuclear spins as qubits were investigated with bulk $^{nat}$Si and isotopically enriched IKZ $^{29}$Si at Stanford University, USA and Keio University, Japan.[69] The decoupling pulse sequences were used to extend the $^{29}$Si coherence time up to 20 s and beyond at room temperature.[69]

Confirmation of long-enough coherence times in various electron and nuclear spins in silicon encouraged researchers to develop qubit initialization schemes. While initialization of the electron spins can be achieved by reducing the sample temperature and increasing the externally applied magnetic field to make the thermal energy much smaller than the Zeeman split electronic levels,[38, 70] initialization of the nuclear spins is not straightforward. This is because of the extremely small gyromagnetic ratios of the nuclear spins that do not allow a large enough Zeeman energy separation with respect to the thermal energy, even with an externally applied magnetic field ~10 T at a temperature of ~1 K. Therefore, it was necessary to develop a method to polarize the nuclear spins in silicon using reasonable experimental conditions. Again, isotope engineering was proven useful for such purposes. The international collaboration led by Simon Fraser University developed an all-optical NMR of $^{31}$P using the Avogadro bulk $^{28}$Si samples.[71] When the inhomogeneous broadening inherent in natural Si was eliminated by using enriched $^{28}$Si, the photoluminescence excitation (PLE) spectrum of the ensemble of phosphorus bound excitons revealed a well-resolved hyperfine splitting resulting form the $^{31}$P nuclear spin (Fig. 3).[71-73] Furthermore, introducing one more laser irradiation at one of the phosphorus PLE peak frequencies, corresponding to the $^{31}$P nuclear spin up or down, led to extremely fast (~0.1 s) electron and nuclear polarizations of 90% and 76%, respectively.[74] This fast initialization and nuclear spin readout technique was employed to demonstrate the long $^{31}$P nuclear coherence times reported in Refs. [63, 64]. By using isotopically enriched $^{28}$Si device structures, it was demonstrated that all-electrical $^{31}$P nuclear polarization was possible.[75] Inhomogeneous broadening of the bismuth bound exciton photoluminescence peaks were also dominated by variations in the masses of the silicon isotopes.[76] $^{28}$Si enrichment allowed for the $^{209}$Bi nuclear spin peaks to be clearly observed.[77]

Achieving a large polarization for the $^{29}$Si nuclear spins in silicon was challenging because, unlike donor nuclear spins, there is no electron spin that couples strongly to the $^{29}$Si at the substitutional lattice sites. One approach taken by the international collaboration led by Keio University was to enrich the phosphorus doped single silicon crystal with $^{29}$Si to reduce the average distance between the phosphorus electrons and the $^{29}$Si nuclear spins.[78-81] In this phenomena, referred to as dynamical nuclear polarization, selective excitation of the symmetry prohibited electron spin resonance transitions in the phosphorus donor coupling to the nearby $^{29}$Si, leading to preferential relaxation to a certain $^{29}$Si nuclear spin state (solid effect). This nuclear spin polarization around the phosphorus atom diffused through the crystal via nuclear spin diffusion ($^{29}$Si-$^{29}$Si flip-flops). However, this process is slow (>1 h) and the highest $^{29}$Si polarization achieved was only ~4%.[81] An efficient method would be to prepare electron spins that couple more directly to the $^{29}$Si nuclear spins, similar to the way that an electron bound to a phosphorus atom couples strongly to the $^{31}$P nuclear spin. To achieve this condition, the group led by Keio University employed vacancy-oxygen complexes in silicon that had $^{29}$Si nearest neighbors.[82-84] This defect is unique as it can be excited into the spin triplet state to achieve large electron spin polarization that can be transferred to the nearby $^{29}$Si nuclear spins. This allowed a very large, most likely >90%, polarization of the $^{29}$Si nuclear spins.[83]

Taking the advantages of the long coherence time and the absence of inhomogeneous broadening in the phosphorus electron spin resonance (ESR) and electron nuclear double resonance (ENDOR) lines for the phosphorus donors in isotopically enriched IKZ bulk $^{28}$Si, the international collaboration led by John Morton (formally at University of Oxford and presently at University College London) succeeded in a series of proof-of-concept two-qubit experiments. Here, one qubit was the electron spin of the phosphorus atom and the other was the $^{31}$P nuclear spin. By sharpening the electron ESR and $^{31}$P ENDOR transitions in $^{28}$Si, the quantum manipulation fidelity of the ensemble of phosphorus donor states was improved significantly. Additionally, eliminating the $^{29}$Si nuclear spins from the matrix provided enough coherence time to complete the operation and readout procedures, such as the coherent transfer of the quantum states,[38] and create and detect the two-qubit entanglement[70] between the phosphorus electron and $^{31}$P nuclear spins. The same team also succeeded in employing an ensemble of phosphorus electron spins to store and retrieve multiple microwave pulses, demonstrating the basic operation of the so-called quantum holographic memory[85] and the geometric phase gate operation of the phosphorus electron spins for fault-tolerant quantum computation.[86] They also performed fundamental experiments testing the nature of the quantum superposition states.[87] In parallel, the Keio-UCL collaboration demonstrated coherent transfer between the electron spin states and $^{29}$Si nuclear spin states for two-qubit operations.[83] These accomplishments, obtained with ensembles of donor electron spins, donor nuclear spins, and $^{29}$Si nuclear spins in silicon formed the foundation of single bit quantum information processing in silicon, which will be discussed in the next section.

The properties of ensembles containing boron acceptors were studied using IKZ bulk $^{28}$Si single crystals. It is well known that the electron spin resonance spectra of acceptors are broadened by the existence of the random strains that lift the degeneracy of the light and heavy hole bands from one acceptor to another by different amounts. However, the main source of random strain has not been identified. The group led by Keio University and TUM identified that the source of strain was the random distribution of the three stable isotopes in $^{nat}$Si. They observed significant narrowing of the boron ESR lines in $^{28}$Si.[88, 89] However, the coherence time of the holes bound to the boron acceptors were limited by the short hole relaxation time ($T_1$), even at 3 K. Further cooling is needed to extend the $T_1$ and $T_2$ for the application of boron acceptors in quantum information processing. Additionally, the properties of the donors have been studied in isotopically enriched, $^{28}$Si device structures. The group led by Berkeley investigated the spin-dependent scattering in $^{28}$Si field-effect transistors.[90] The formation of $^{29}$Si nuclear spin wires was also attempted by templating a silicon substrate,[91] followed by forcing the alignment of MBE-deposited Si atoms.[92] However, a $^{29}$Si device is yet to be realized.

### D. SINGLE SPIN QUBITS IN SILICON

The most important steps towards realizing a silicon-based quantum computer are to prepare, operate, and readout a single charge (or double quantum dot) qubit or single electron and/or nuclear spin qubit placed in a silicon matrix. While, truly single qubit operations in silicon have been reported in the past few years,[93-96] it was only this year that the isotope engineering of silicon was proven essential, even for single qubits in silicon.[97, 98] For a single qubit, such as a single donor, a single quantum dot or a single double-quantum dot, a variety of foreign materials and structures including gate insulators, metal contacts, and SETs for readouts should be placed nearby. Therefore, unlike the ensemble of isolated donors described in the previous section whose coherence was limited by the presence of $^{29}$Si nuclear spins in $^{nat}$Si, the common assumption was that purifying the isotope to eliminate $^{29}$Si would not be effective in extending the coherence time because the major decoherence sources would come from foreign objects rather than the $^{29}$Si nuclear spin fluctuations. Despite such skepticism, the "let's try and see" approach employed by experimentalists has shown that isotope enrichment is essential to successfully fabricate single donor spin qubits[97] and single quantum dot qubits[98] in silicon.

Figure 4 shows a scanning electron microscope (SEM) image of the single phosphorus device developed by the group led by Andera Morello and Andrew Dzurak at the University of New South Wales (UNSW). This device allowed, not only initialization, operation, and readout of the quantum information stored in a single electron qubit and the single $^{31}$P nuclear qubit associated with the single phosphorus atom, but also coherent transfer between the electron and nuclear spin qubits. The device structure was essentially the same as the one developed by the same group using a $^{nat}$Si substrate, demonstrating single electron spin[94] and nuclear spin[95] qubits. Figure 5(a) and (b) compares the Rabi oscillations of the single phosphorus qubits in $^{nat}$Si[94] and $^{28}$Si, respectively.[97] The single electron spin qubit placed near the surface had a much longer coherence ($T_2^*$) when isotopically enriched $^{28}$Si was employed. In fact, every aspect of the qubit properties were significantly improved in the $^{28}$Si wafer (summarized in Table I).

Isotope engineering was also proven to be highly beneficial for the single electron spin quantum dot qubits in the $^{28}$Si samples fabricated and measured by the same group at UNSW.[98] Figure 6 shows the device structure and the Rabi oscillations of the electrons. Table I summarizes the coherence times that are orders of the magnitude longer than other quantum dot qubits. Moreover, the electron spin resonance linewidth of the quantum dot qubit was only 2.4 kHz and its resonant frequency could easily be shifted by more than 3000 times the linewidth by tuning the gate voltage. This ensured the selectivity to specific qubits integrated in large-scale systems, which could be fabricated using the standard silicon complementary metal-oxide-semiconductor (CMOS) technologies.

Developments in the design of quantum mechanically coupled, multiple phosphorus qubits and quantum dots are required for future large-scale integration. The two experiments described in this section showed that using isotope engineering to eliminate the background nuclear spins is crucial to extend the coherence time to a useful level. It was also shown in Fig. 2 of Ref. 99 that the isotope enrichment is needed to improve the gate fidelity. Overall, single qubits in $^{28}$Si have greatly surpassed the properties of the other single qubits tested in semiconductor systems.

### III. DIAMOND QUANTUM SENSING

#### A. DIAMOND QUANTUM SENSING

Diamond has been identified as one of the ideal matrixes for use in solid-state quantum information processing (QIP)[100] when ESR of a single nitrogen-vacancy (NV) center was realized.[101] The group from the University of Stuttgart, Germany performed a series of pioneering works on the qubit operations of single NV centers at room temperature.[7, 9, 102, 103] Their ground-breaking works were extended to a variety of applications of single NV centers in QIP.[104-109] In parallel, studies on how to use quantum phenomena in metrology and sensing emerged, aiming to reach sensitivities beyond the limit imposed by classical physics.[110, 111] While discussions continue on the ultimate limit of quantum sensing,[112-116] the use of single electron spins bound to single NV centers in diamond has been proposed.[11, 12, 14] An example of quantum sensing by the NV centers is shown in Fig. 7, demonstrating the basic configuration used to probe the magnetic field that arose from the single nuclear spin of an atom, which is one of the constituents of a molecule. A single NV (blue arrow) embedded in a diamond substrate was used as a quantum sensor to probe the field coming from a molecule placed on the surface (Fig. 7(a)), while a single NV placed at the tip of a cantilever was used as a quantum sensor (Fig. 7(b)). To realize such measurements, two criteria must be fulfilled: i) the electron spin of each NV must be tilted by 90 ° with respect to the direction of the externally applied magnetic field (B) so that it precesses around the B-field axis. The change in the precession rate is measured to probe the magnetic field arising from the nuclear spins of the molecule, thus, longer precession times have better sensitivities. The precession hold time decays exponentially and the coherence time ($T_2^*$) for the DC measurements and the $T_2$ for the AC measurements are the relevant figures-of-merit because the minimum magnetic field required is proportional to $1/\sqrt{T_2}$. ii) Because the magnetic field strength decreases by $1/d^3$, where *d* is distance, the distance between the target nuclear spins and the NV electron sensor must be reduced, preferably to the order of a few nanometers, to detect single nuclear spins. The NV center has advantages over other quantum system because its single electron spin, $T_2$, is long and can be measured at room temperature. Because a NV electron is spatially confined within 1 nm, it can be placed within a few nanometers from the target. Therefore, research leading to the application of NV centers for magnetic sensing has become very active recently[117-128] and the use of such sensors in solid-state physics research is emerging.[129-131] Because the electronic levels of NVs are affected by temperature[132] and other external perturbations, NVs can be used to sense temperature,[133-135] electric fields,[136] pressure,[137] and mechanical motions.[138] Moreover, diamond is biocompatible and harmlessness, allowing for in-vivo quantum sensing in biology and medicine.[139-141]

### B. DIAMOND ISOTOPE ENGINEERING FOR QUANTUM SENSING

Pioneering works on the isotope engineering of bulk diamonds have been performed at the National Institute for Research in Inorganic Materials (currently called the National Institute of Materials Science) in Japan[142] and at the General Electric Company, USA.[143] The isotope effect on the band gap was investigated at Waseda University using $^{13}$C isotopically enriched CVD diamond.[144] The National Institute of Advanced Industrial Science and Technology (AIST) in Japan pioneered an isotope superlattice consisting of alternating thin layers of $^{12}$C and $^{13}$C.[145-147] $^{12}$C and $^{13}$C enriched methane gases used in CVD at AIST were obtained from Cambridge Isotope Laboratories, Inc. Similar to silicon, the coherence time of the NV centers was studied in isotopically controlled diamond as a function of the background $^{13}$C nuclear spin concentration.[148] It was extended to ~2 ms at room temperature by isotopic enrichment with nuclear-spin-free $^{12}$C.[15] Here, the isotopically enriched diamond grown by Element Six was employed. The long coherence time was maintained in 100-nm-thick, isotopically enriched $^{12}$C

CVD single crystalline films where NV centers were introduced by nitrogen doping during growth.[149] In the isotopically enriched, $^{12}$C CVD polycrystalline films, NV centers were introduced by nitrogen doping during growth.[150-152] In the isotopically enriched $^{12}$C CVD films, NV centers were introduced by nitrogen ion implantation and post implantation annealing.[153, 154] Thus, the remaining challenge for the application of quantum sensors is placing negatively charged NV$^-$ centers as close as possible to the diamond surface. If the NV$^-$ center loses its paramagnetic electron spin, it becomes a neutral NV$^0$ center and cannot function as a quantum sensor. This happens near the surface of the film (within 5 nm) because of the existence of surface states that can change significantly with the crystalline quality of the surface and variations in surface terminating species.[155-157] To improve the crystalline quality, CVD isotopically enriched $^{12}$C diamond films were used. Shallow NV$^-$ centers can be introduced into isotopically enriched CVD $^{12}$C diamond films either by nitrogen doping during CVD,[149, 158] which is sometimes followed by electron beam irradiation to create vacancies,[159] or by ion implantation and annealing.[160] Figure 8 compares the typical NV$^-$ electron $T_2$ values, measured by the Hahn echo sequence at room temperature. Figure 8(a) shows an isotopically enriched $^{12}$C bulk crystal.[15] Figure 8(b) shows a 100-nm-thick isotopically enriched $^{12}$C film[149] and Fig. 8(c) displays a 5-nm-thick isotopically enriched $^{12}$C film.[158] For the 100-nm-thick film, $T_2 \sim 1.7$ ms (Fig. 8(b)). This was the same as the bulk value of $T_2 \sim 1.7$ ms (Fig 8(a)). $T_2 \sim 45$ μs in the 5-nm-thick film, which was shortened significantly by decoherence, most likely caused by the surface states. Making use of the long NV$^-$ electron $T_2$, extended by $^{12}$C enrichment, a single NV$^-$ situated in the deep region (away from the surface) has been used to detect a single $^{13}$C nuclear spin situated either next to the NV$^-$[161] or far away from it, but still in the same diamond crystal.[162] In a similar manner, a shallow NV⁻ placed near the surface should be able to detect nuclear fields that originate from the nuclear spins on the surface of the diamond. In this regard, the $T_2$~45 μs in the 5-nm-thick film was long enough (at least theoretically) to detect a magnetic field from a single proton nuclear spin.[158] There have been two types of isotopically enriched, $^{12}$C diamond CVD films containing shallow NV⁻ centers that have succeeded in detecting the magnetic field from a small ensemble of proton nuclear spins; one was produced by UC Santa Barbara (UCSB)[159] and the other was fabricated by the Keio University–AIST collaboration in Japan.[158] While the UCSB sample was employed successfully by the IBM Almaden Research Center to detect the NMR of the proton spins confined in 24 nm³ of polymethyl methacrylate (PMMA), placed directly on the diamond surface,[163] the Keio-AIST sample succeeded in (along with the Keio-AIST-ETH collaboration) detecting ~6,000 nuclear spins in an immulsion oil placed on the surface of diamond.[158] The nano-scale NMRs of the protons placed on the diamond surface were also demonstrated using shallow NV⁻ centers created in natural diamond.[164, 165] Because the coherence time of the shallow NV⁻ is currently limited by some sorts of defects near the surface rather than by the presence of $^{13}$C nuclear spins, advantage of isotopically enriched $^{12}$C is limited to the absence of oscillations in the echo decay (ESEEM) of the NV electron precession arising from the precession of the nearby $^{13}$C nuclear spins. ESEEM is an additional noise that occurs when the nuclear spins on the surface are the targets for detection. However, the coherence times ($T_2$) of the shallow NVs in both natural and $^{12}$C diamonds are limited by the fluctuations that arise from the surface electronic states. At present, efforts are focused on identifying and understanding the nature of the surface electron spins using single shallow NV⁻ centers as a sensor.[166-168] However, for the noise spectra, such as the intensity of the detected AC magnetic fields vs. AC frequency, there is a disagreement between publications, indicating that the near-surface spin states are different for samples that have been prepared differently. Advancements in the materials science of diamond thin film growth and surface preparation are needed to understand why these differences occur. One interesting direction explored recently was investigating the surface orientation rather than the (100) direction, which is usually studied. Additionally, recently the (111) surface has been extensively studied.[169-170, 171] Other defects in diamond, such as silicon vacancy complexes, may be of interest for sensing in the future. Such defects have been studied recently using isotopically enriched $^{12}$C diamond.[172]

## IV. SUMMARY AND OUTLOOK

State-of-the-art materials science and quantum information experiments involving isotope engineering have been reviewed for silicon and diamond. The role of isotope engineering in silicon quantum computing has been shown to be crucial. The implementation of isotope engineering in todays advanced silicon material technologies is fully developed. However, the availability of isotopically enriched $^{28}$Si in industrially adopted forms, such as silane and trichlorosilane, are limited. The advantages of the isotope engineering of diamond for quantum sensing applications have been shown. However, while the availability of $^{12}$C in various forms is not a problem, further improvements in increasing the crystallinity and understanding of the surface states in diamond are needed in order to take full advantage of the possibilities of isotopic engineering as already demonstrated in Si.[172]

**Acknowledgments**

We thank Mike Thewalt, John Morton, Steve Lyon, Andrea Morello, Andrew Dzurak, Martin Brandt, Thomas Schenkel, Christian Degen, Fedor Jelezko, Tokuyuki Teraji, Junichi Isoya, and Junko Hayase for their helpful discussions. The research was supported in part by a Grant-in-Aid for scientific research by MEXT and by the JSPS Core-to-Core Program.

**Table I Comparison of the fidelities and coherence times of phosphorus-donor and quantum-dot qubits in naturally available silicon and isotopically enriched $^{28}$Si epilayers.**

|  | Single phosphorus qubit in $^{nat}$Si[94, 95] | Single phosphorus qubit in $^{28}$Si[97] | Single quantum dot qubit in $^{28}$Si[98] |
|---|---|---|---|
| Electron spin measurement fidelity ($F_M$) | 77% | 97% | 92% |
| Electron spin control fidelity ($F_C$) | 57% | >99.5% | >99.5 |
| Electron $T_2^*$ | 0.0055 μs | 140 μs | 120 μs |
| Electron $T_2$ determined by Hahn echo | 0.21 ms | 0.95 ms | 1.2 ms |
| Electron $T_2$ extended by CPMG | -- | 559 ms | 28 ms |
| Electrically neutral donor $^{31}$P nuclear $T_2^*$ | 0.84 ms | 0.57 ms | -- |
| Electrically positive donor $^{31}$P nuclear $T_2^*$ | 3.3 ms | 600 ms | -- |
| Electrically neutral donor $^{31}$P nuclear $T_2$ | 3.5 ms | 20.4 ms | -- |
| Electrically positive donor $^{31}$P nuclear $T_2$ | 60 ms | 1.75 s | -- |

**Figure captions**

Fig. 1. Two examples of silicon quantum computer schemes using isotope engineering. (a) Phosphorus donor nuclear spin qubits[3] and (b) $^{29}$Si nuclear spin qubits.[6]

Fig. 2. Schematics of (a) quantum mechanically different phosphorus donors placed in $^{nat}$Si and (b) quantum mechanically identical phosphorus donors placed in isotopically enriched $^{28}$Si.

Fig. 3. Comparison of the photoluminescence excitation spectra for an ensemble of phosphorus-bound excitons in $^{nat}$Si and isotopically enriched $^{28}$Si. The spectrum of $^{nat}$Si is significantly broader because of the random distribution of the three stable isotopes, while that of $^{28}$Si has sharp doublet features, corresponding to the hyperfine splittings due to the $^{31}$P nuclear spins.[71]

Fig. 4. SEM image of the single phosphorus spin qubit device developed at UNSW. The red spin indicates where the single phosphorus qubit is placed.[97]

Fig. 5. Comparison of the Rabi oscillations for the single phosphorus qubit placed in (a) $^{nat}$Si[94] and (b) isotopically enriched $^{28}$Si.[97]

Fig. 6. (a) SEM image of a gate-defined single electron quantum dot qubit in silicon. (b) Rabi oscillations for a single electron spin confined in a quantum dot.[98]

Fig. 7. Sensing the magnetic field that arose from the nuclear spins of a molecule using (a) a single NV center embedded in a diamond substrate and (b) a single NV center embedded in the tip of a diamond cantilever.

Fig. 8. Comparison of the typical NV⁻ electron $T_2$ values measured by the Hahn echo sequence at room temperature in (a) an isotopically enriched $^{12}$C bulk crystal,[15] (b) a 100-nm-thick isotopically enriched $^{12}$C film[149] and (c) a 5-nm-thick isotopically enriched $^{12}$C film.[158]

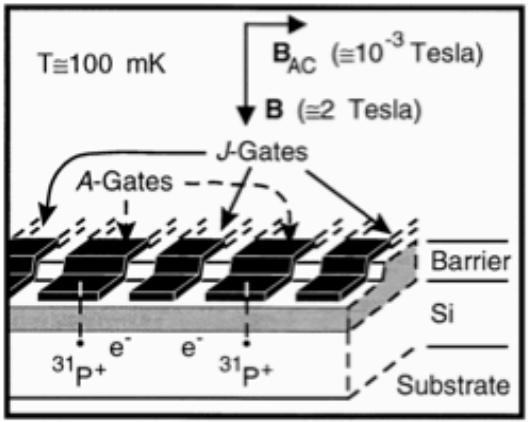 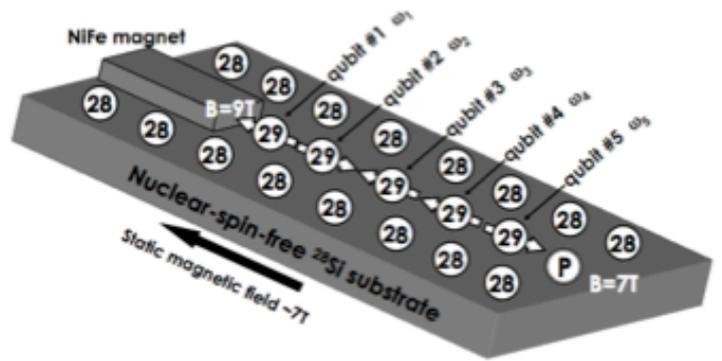

Fig. 1

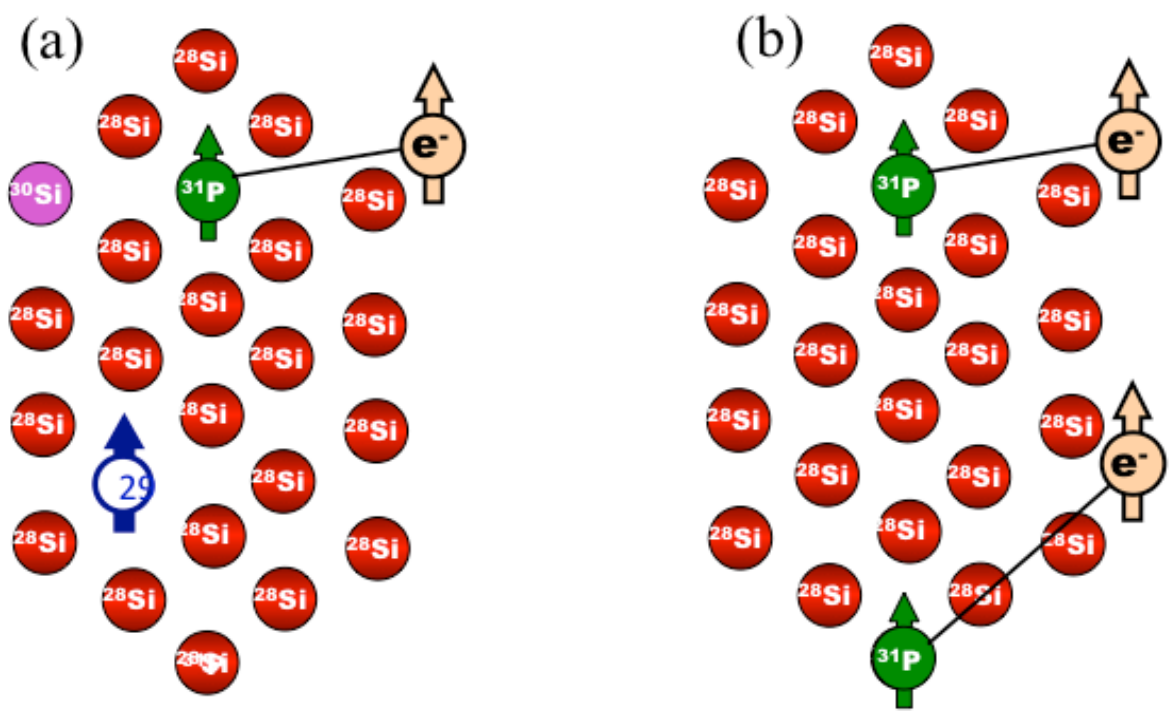

Fig. 2

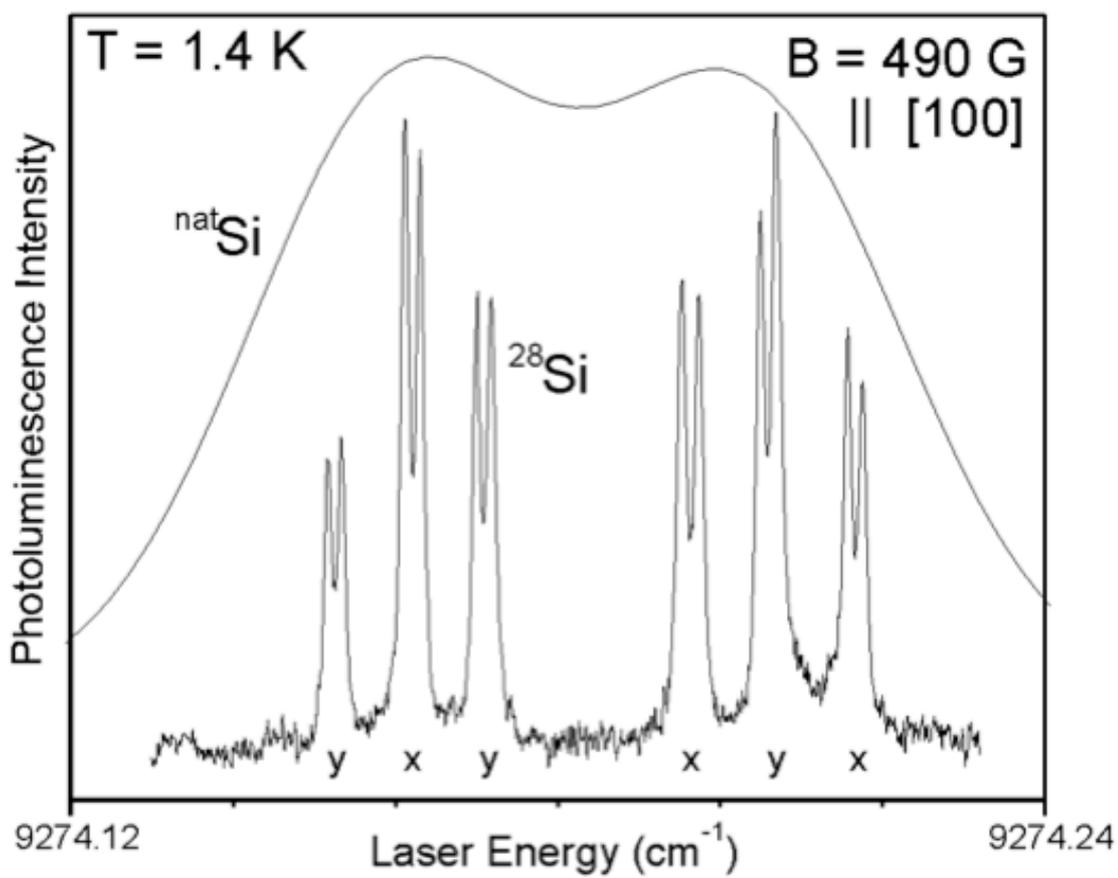

Fig. 3

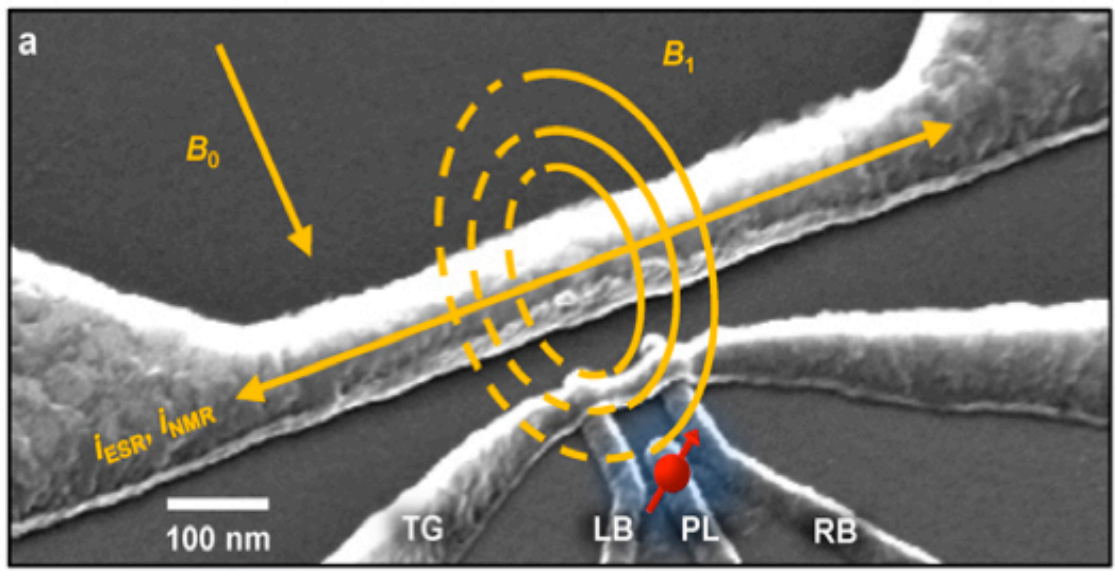

Fig. 4

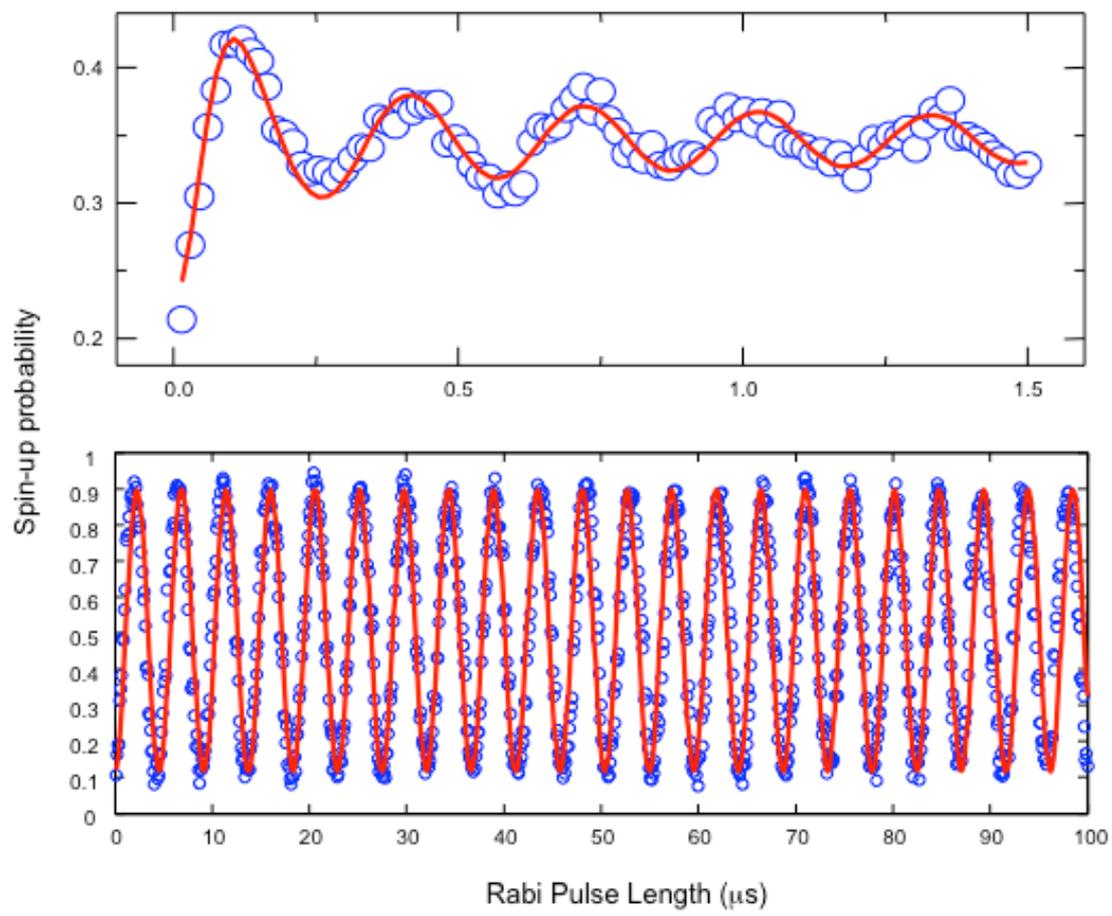

Fig. 5

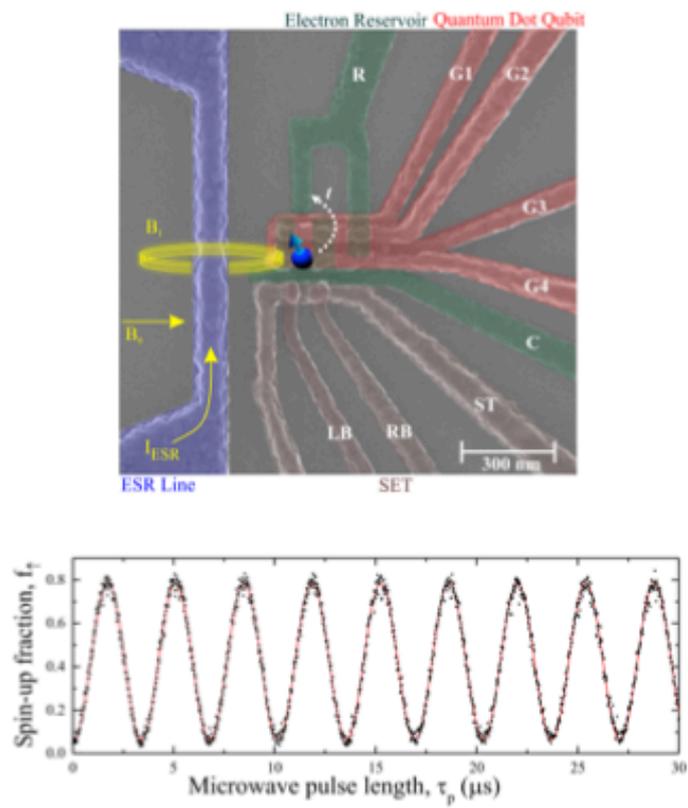

Fig. 6

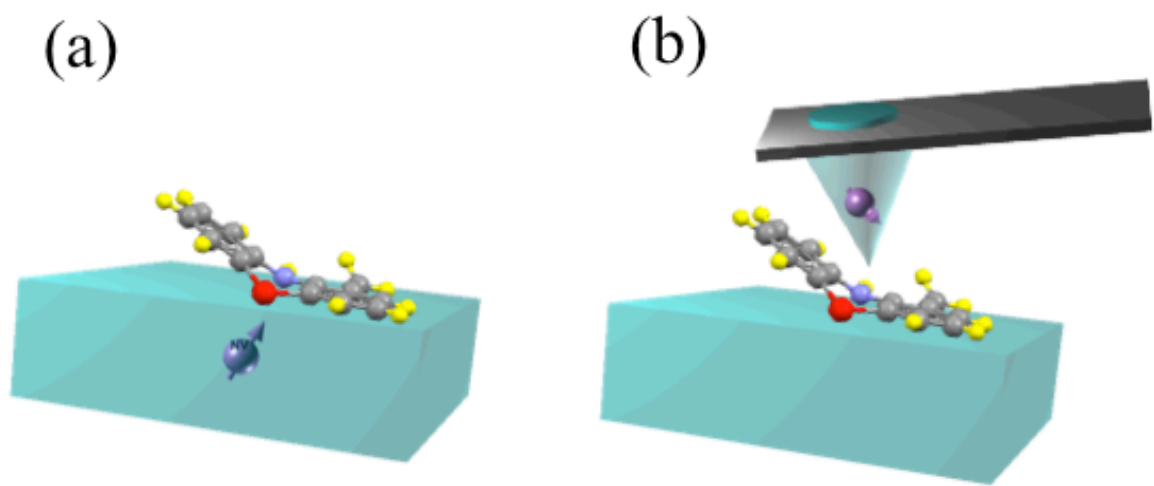

Fig. 7

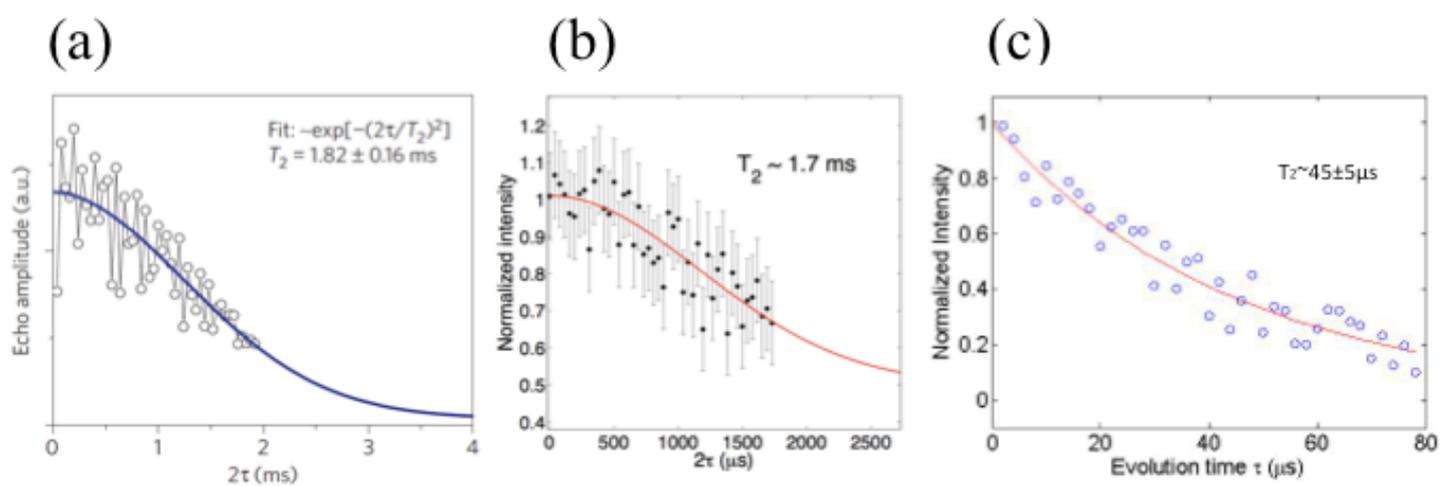

Fig. 8